\documentclass[12pt]{iopart}

\usepackage{iopams}
\usepackage{graphicx}
\begin{document}

\title[Cavity losses for the dissipative Jaynes-Cummings Hamiltonian beyond RWA]{Cavity losses for the dissipative Jaynes-Cummings Hamiltonian beyond
Rotating Wave Approximation}

\author{M. Scala, B. Militello, A. Messina}

\address{MIUR and Dipartimento di Scienze Fisiche ed
Astronomiche dell'Universit\`{a} di Palermo, via Archirafi 36,
I-90123 Palermo, Italy} \ead{matteo.scala@fisica.unipa.it}

\author{S. Maniscalco, J. Piilo and K.-A. Suominen}

\address{Department of Physics, University of Turku, FI-20014
Turun Yliopisto, Finland}

\begin{abstract}
 A microscopic derivation of the master equation for the
 Jaynes-Cummings model with cavity losses is given, taking into
 account the terms in the dissipator which vary with frequencies
 of the order of the vacuum Rabi frequency. Our approach allows to
 single out physical contexts wherein the usual phenomenological
 dissipator turns out to be fully justified and constitutes an extension
 of our previous analysis [Scala M. {\em et al.} 2007 Phys. Rev. A {\bf 75}, 013811],
 where a microscopic derivation was given in the framework of the Rotating Wave
 Approximation.
\end{abstract}



\maketitle

\section{Introduction}
Since its first appearance~\cite{JC}, the Jaynes-Cummings (JC)
model has been extensively used to study radiation-matter
interaction~\cite{shore_review} in contexts like cavity quantum
electrodynamics (CQED)~\cite{haroche,harocheRMP,Walther} and ion
traps~\cite{wineland}. In a CQED scenario~\cite{haroche}, for
instance, the model describes the one-photon interaction between a
two-level atom and a quantized normal mode of the electromagnetic
cavity field, while in ion traps~\cite{wineland} it describes the
interaction between a two-level atom and a vibrational mode of its
center of mass. In both contexts, the quantum effects predicted by
the model, such as Rabi oscillations and collapses and revivals of
the atomic inversion operators, have been experimentally observed.

Concentrating on CQED, it is of central importance to describe
cavity losses, due for example to dissipative effects in the
cavity mirrors. In Ref.~\cite{Scala} we addressed this problem and
provided a microscopic derivation for the master equation for the
JC model neglecting spontaneous emission and supposing that the
cavity was coupled to a bosonic reservoir. The derived master
equation was then compared to the phenomenological master equation
used in the literature~\cite{haroche} and some physical examples
were given to show that the two models describe very different
decay mechanisms. Indeed, while in the phenomenological model it
is the cavity only which directly decays and the atom loses energy
through its coupling with the cavity mode, in our model it is the
atom-cavity system as a whole which leaks, although only the
cavity is coupled to an external environment.

Our approach in Ref.~\cite{Scala} relied on the rotating wave
approximation (RWA) performed on the dissipator, as done in the
usual formalism of master equations~\cite{petruccionebook}. In
fact it is possible to perform RWA in our case only when the
vacuum Rabi frequency is much larger than the largest decay rate
involved, i.e., in the strong coupling regime. Since the
phenomenological model is claimed to be valid also in regimes
wherein the vacuum Rabi frequency and the cavity decay rate are of
the same order, the comparison between it and the microscopic
model will be complete only when a microscopic derivation is given
which takes into account in the dissipator all the terms
oscillating at frequencies of the order of the vacuum Rabi
frequency.

This is what we are going to do in this paper. Our present
investigation, based on a deeper analysis of the terms neglected
in the so-called dressed-state approximation performed on the
phenomenological model~\cite{haroche}, shows that the microscopic
and the phenomenological model can be much closer than one might
think in the context of our previous RWA model. In particular we
will see that the two models may coincide if the spectrum of the
environment is flat, even for small nonzero temperatures.

The paper is structured as follows. In Sec.~2 we review the JC
model, the usual phenomenological way to describe cavity losses
and our microscopic model derived in Ref.~\cite{Scala}, discussing
its validity and limitations. In Sec.~3 we derive the microscopic
model beyond RWA and in Sec.~4 the latter is compared to the
phenomenological model, singling out the conditions under which
the two models coincide, both exactly and approximately. Finally
in Sec.~5 some conclusive remarks are given.

\section{The Jaynes-Cummings model: the phenomenological description of cavity losses}

In Ref.~\cite{Scala} we addressed the problem of the description
of the dissipation and decoherence processes in a two-level
atom-cavity system, neglecting spontaneous emission, and taking
into account the coupling between the cavity and an external
bosonic environment. Denoting by $\left|g\right>$ and
$\left|e\right>$ the atomic ground and excited states
respectively, and calling $\omega_0$ the atomic Bohr frequency,
the atom-cavity interaction, in the RWA and in units of $\hbar$,
at resonance, is described by the Jaynes-Cummings
Hamiltonian~\cite{JC}:
\begin{eqnarray}\label{JCM}
 H_{JC}=\frac{\,\omega_0}{2}\sigma_z+\omega_0\,a^\dag
 a+\Omega\left(a\sigma_++a^\dag\sigma_-\right).
\end{eqnarray}
Here $a^\dag$ ($a$) denotes the creation (annihilation) operator
of the electromagnetic mode, and the atomic degrees of freedom are
described by $\sigma_-=\left|g\right>\left<e\right|$,
$\sigma_+=\left|e\right>\left<g\right|$, and $\sigma_z =
\left|e\rangle \langle e \right| - \left|g \rangle \langle g
\right|$.The eigenstates and eigenvalues of $H_{JC}$
are~\cite{shore_review}:
\begin{eqnarray}\label{eigenstates}
 &&\left|E_{N,\pm}\right>=\frac{1}{\sqrt{2}}\left(\left|N,g\right>\pm
 \left|N-1,e\right>\right),\nonumber\\
 \nonumber\\
 &&E_{N,\pm}=\left(N-\frac{1}{2}\right)\omega_0\pm\Omega\sqrt{N},
\end{eqnarray}
except for the ground state which is given by:
\begin{eqnarray}\label{groundeigenstate}
 \left|E_0\right>=\left|0,g\right>, \hspace{1cm}
 E_0=-\frac{\omega_0}{2}\,,
\end{eqnarray}
where $\vert N, i \rangle = \vert N \rangle \vert i \rangle$, with
$i=e,g$, indicates the tensor product of the Fock state $\vert N
\rangle$, with $a^\dag a \vert N \rangle=N\vert N \rangle$, and
the electronic state $\vert i \rangle$.

In the usual treatment of the problem of cavity losses one derives
microscopically the master equation for the cavity only, when it
is coupled to an environment at temperature $T$~\cite{cohen_libro}
and then assumes that the atom inside the cavity causes only a
change in the Hamiltonian governing the unitary part of the
dynamics~\cite{haroche}. In this way the following
phenomenological master equation for the density operator of the
atom-cavity system is assumed to be valid:
\begin{eqnarray}\label{phenME}
 \dot{\rho}&=&-i\left[H_{JC},\rho\right]+\gamma \left[ n(\omega_0)+1 \right]  \left[ a\rho
 a^\dag - \frac{1}{2}\left(a^\dag a \rho+\rho a^\dag
a \right)\right]\nonumber\\
\nonumber\\
&+& \gamma n(\omega_0)  \left[ a^\dag\rho
 a-\frac{1}{2}\left(a a^\dag \rho+\rho a
a^\dag\right)\right],
\end{eqnarray}
where $n(\omega_0)$ is the average number of quanta of the
reservoir in the quantized mode of frequency $\omega_0$, and
$\gamma$ is the rate of loss of cavity photons. For a reservoir at
zero temperature, Eq.~(\ref{phenME}) becomes:
\begin{eqnarray}\label{oldME}
 \dot{\rho}=-i\left[H_{JC},\rho\right]+\gamma\left( a\rho
 a^\dag-\frac{1}{2}a^\dag a \rho -\frac{1}{2} \rho a^\dag
a\right).
\end{eqnarray}

Equations~(\ref{phenME})~and~(\ref{oldME}) are the
phenomenological models that are very often used in the literature
on cavity
QED~\cite{agar1,agar2,agar3,barnett,briegel,puribook,harocheRMP}.

In Ref.~\cite{Scala} we investigated the legitimacy of
Eq.~(\ref{phenME}). By applying the general formalism from the
theory of open quantum systems~\cite{petruccionebook}, we
microscopically derived the master equation for the system under
scrutiny assuming from the very beginning that the Hamiltonian of
the system was the JC model.

Stated another way, the general theory of master equations in the
Born-Markov approximation claims that the decay channels are
described by jump operators which are transition operators between
eigenstates of the Hamiltonian of the system, with coefficients
depending on the details of the coupling between the system and
the environment. This feature of the theory is quite general and
holds even when the system is bipartite and only one of  its two
parts is coupled to an external environment.

By applying this formalism to the JC model, we found~\cite{Scala}
that, although cavity only is coupled to an external environment,
it is instead the atom-cavity system as a whole which decays, the
irreversible transitions giving rise to dissipation and
decoherence through jumps between dressed states. We then
investigated the time evolution of the atom-cavity system
predicted by the phenomenological model and by our microscopic
model to clarify the different physical mechanisms underlying the
two different models.

We stress that this approach includes the so-called rotating wave
approximation (RWA)~\footnote{One should avoid confusion between
the RWA performed on the Hamiltonian in the JC model and the RWA
performed on the dissipator. In the following we will always refer
to RWA as to the operation of neglecting the oscillating terms in
the dissipator in the Interaction Picture when deriving the master
equation.} on the dissipator of the master equation and is at the
origin of its limited range of applicability expressible
as~\cite{Scala}:
\begin{eqnarray}\label{RWA_JC}
 2\Omega\gg\gamma_{\rm max}.
\end{eqnarray}

In other words the RWA in Ref.~\cite{Scala} is done over
timescales of the order of the inverse of the Rabi frequency
$2\Omega$ of the atom-cavity system, and the microscopic master
equation is valid only when the decay occurs on a timescale which
is much longer than the Rabi period. On the other hand the
phenomenological model described by
Eqs.~(\ref{phenME})~and~(\ref{oldME}) uses a dissipator on which
the RWA has been done over frequencies which are of the order of
$\omega_0$, which is much larger than $2\Omega$: therefore the
phenomenological model is claimed to be valid in a much larger
range of values of the decay rate, since in this model one must
have
\begin{eqnarray}\label{RWA_optical}
 \omega_0\gg\gamma.
\end{eqnarray}

In the following we investigate how the master equation for the JC
model is changed when its microscopic derivation does not neglect
those terms in the dissipator oscillating at frequencies of the
order of the Rabi frequency.

\section{The microscopic master equation for the JC model beyond the RWA}
In Ref.~\cite{Scala} we have assumed the validity of the RWA on
timescales of the order of the period of the Rabi oscillations.
Now we instead take into account the terms in the dissipator
oscillating at frequencies proportional to $\Omega$, in order to
obtain a microscopic model suitable for a comparison with the
phenomenological model in all damping regimes.

What we are going to do now is to skip the step corresponding to
RWA in the general formalism to get a master equation under the
Born and Markov approximations and to see what we can say about
the master equation for the JC model starting from the following
model for the total closed system:
\begin{eqnarray}\label{modello}
 H_S&=&H_{JC},\;\;\;H_E=\sum_k\omega_k b_k^\dag b_k, \nonumber\\
            \nonumber\\
            H_{\rm int}&=&\left(a+a^\dag\right)\sum_kg_k\left(b_k+b_k^\dag\right),
\end{eqnarray}
which describes an atom-cavity system wherein the cavity is
coupled to a bosonic environment through the interaction
hamiltonian $H_{\rm int}$. This model is equivalent to that used
in Ref.~\cite{cohen_libro}, since the additional terms rapidly
oscillating at frequencies of the order of $\omega_0$ are washed
out in the total and partial RWAs performed in the following.

From the interaction Hamiltonian $H_{\rm int}$ one gets the jump
operators, according to the recipe given in
Ref.~\cite{petruccionebook}:
\begin{eqnarray}\label{aomegabis}
 &A&\!\!\!\left(\!E_{N'\!,\,l}\!-\!E_{N,\,m}\right)=\left|E_{N,\,m}\right>\left<E_{N,\,m}\right|\!\left(a+a^\dag\right)\!
                        \left|E_{N'\!,\,l}\right>\left<E_{N'\!,\,l}\right| \nonumber \\
                        &=&\frac{1}{2}\delta_{N,N'-1}\!\left(\!\sqrt{N+1}+lm\sqrt{N}\right)\left|E_{N,\,m}\right>\left<E_{N+1,\,l}\right|,
\end{eqnarray}
for $N\ge1$ and
\begin{eqnarray}\label{aomegabisN1}
A\left(E_{1,\pm}-E_0\right)=\frac{1}{\sqrt{2}}\left|E_0\right>\left<E_{1,\pm}\right|,
\end{eqnarray}
for $N=0$. In Eq.~(\ref{aomegabis}) we indicate the states
$\left|E_{N,\pm}\right>$ by $\left|E_{N,\pm 1}\right>$ and the
energy eigenvalues $E_{N,\pm}$ by $E_{N,\pm 1}$. Accordingly $l$
and $m$ take the values $\pm1$. So all the possible jump operators
are transition operators describing jumps between eigenstates
(dressed states) of $H_{JC}$ differing for one excitation only.

We will use the following notation for the jump operators:
\begin{eqnarray}\label{jumpoperators}
  A_{Nlm}=A(E_{N+1,l}-E_{N,m})
\end{eqnarray}
for $N\ge1$ and
\begin{eqnarray}\label{jumpoperators0}
  A_{0l}=A(E_{1,l}-E_0)
\end{eqnarray}
for $N=0$.

Before explicitly inserting these operators in the master
equation, let us examine the structure of the equation for the
reduced density operator of the atom-cavity system, after the Born
and Markov approximations. The equation can be cast in this form:
\begin{eqnarray}\label{born-markov-noRWA-freqpos}
 \frac{d}{dt}\rho(t)=\sum_{\omega,\omega'\ge0}\left\{\mathrm{e}^{i\left(\omega'-\omega\right)
 t}\Gamma(\omega)\left[A(\omega)\rho(t)A^\dag(\omega')-A^\dag(\omega')A(\omega)\rho(t)\right]+\mathrm{h.c.}\right\}\nonumber\\
 \nonumber\\
 +\sum_{\omega,\omega'>0}\left\{\mathrm{e}^{i\left(\omega-\omega'\right)
 t}\Gamma(-\omega)\left[A^\dag(\omega)\rho(t)A(\omega')-A(\omega')A^\dag(\omega)\rho(t)\right]+\mathrm{h.c.}\right\}\nonumber\\
 \nonumber\\
 +\sum_{\omega\ge0,\omega'>0}\left\{\mathrm{e}^{-i\left(\omega'+\omega\right)
 t}\Gamma(\omega)\left[A(\omega)\rho(t)A(\omega')-A(\omega')A(\omega)\rho(t)\right]+\mathrm{h.c.}\right\}\nonumber\\
 \nonumber\\
 +\sum_{\omega>0,\omega'\ge0}\left\{\mathrm{e}^{i\left(\omega'+\omega\right)
 t}\Gamma(-\omega)\!\!\left[A^\dag(\omega)\rho(t)A^\dag(\omega')\!-\!A^\dag(\omega')A^\dag(\omega)\rho(t)\right]\!+\!\mathrm{h.c.}\!\right\}\!,
\end{eqnarray}
where we have explicitly written the minus sign for each negative
Bohr frequency. The coefficients $\Gamma(\omega)$ are complex
functions of the particular Bohr frequency involved and are equal
to:
\begin{eqnarray}\label{Gamma}
 \Gamma(\omega)=\int_{0}^{+\infty}d\tau\,
 \mathrm{e}^{i\omega\tau}\left<E(\tau)E(0)\right>,
\end{eqnarray}
with
$E(\tau)=\sum_kg_k(b_k\mathrm{e}^{-i\omega_k\tau}+b_k^\dag\mathrm{e}^{i\omega_k\tau})$.

All the positive Bohr frequencies involved in the master equation
can be written in the following way:
\begin{eqnarray}\label{bohrfreq}
 \omega_{Nlm}&=&E_{N+1,l}-E_{N,m}=\omega_0+\left(l\sqrt{N+1}-m\sqrt{N}\right)\Omega,
\end{eqnarray}
from which, remembering that $\omega_0\gg\Omega$, it follows that
the first two rows in the master equation are composed of terms
which are either stationary or oscillating at frequencies
proportional to the Rabi frequency, since:
\begin{eqnarray}\label{diffbohrfreq}
 &&\omega_{Nlm}-\omega_{N',kn}=\left(l\sqrt{N+1}-m\sqrt{N}-k\sqrt{N'+1}+n\sqrt{N'}\right)\Omega,
\end{eqnarray}
while the last two rows are composed of terms oscillating at much
larger frequencies, since:
\begin{eqnarray}\label{sumbohrfreq}
 &&\omega_{Nlm}+\omega_{N',kn}=2\omega_0+\left(l\sqrt{N+1}-m\sqrt{N}+k\sqrt{N'+1}-n\sqrt{N'}\right)\Omega.
\end{eqnarray}

Now we will neglect only the last two rows in
Eq.~(\ref{born-markov-noRWA-freqpos}), i.e., all the rapidly
oscillating terms, whereas in Ref.~\cite{Scala} we neglected all
the time-dependent terms, including the slow ones: we will call
{\em Quasi-RWA Master Equation} the resulting equation.

Neglecting these terms in Eq.~(\ref{born-markov-noRWA-freqpos})
and going back to the Schr\"odinger picture, we then get:
\begin{eqnarray}\label{born-markov-noRWA-noFast}
 \frac{d}{dt}\rho(t)&=&-i\left[H_{JC},\rho(t)\right]\nonumber\\
 \nonumber\\
 &+&\sum_{\omega,\omega'\ge0}
 \left\{\Gamma(\omega)\left[A(\omega)\rho(t)A^\dag(\omega')-A^\dag(\omega')A(\omega)\rho(t)\right]+\mathrm{h.c.}\right\}\nonumber\\
 \nonumber\\
 &+&\sum_{\omega,\omega'>0}\left\{\Gamma(-\omega)\left[A^\dag(\omega)\rho(t)A(\omega')-A(\omega')A^\dag(\omega)\rho(t)\right]+\mathrm{h.c.}\right\}.
\end{eqnarray}

Inserting the explicit form of the jump operators for our model,
we finally obtain:
\begin{eqnarray}\label{micros_model}
 \frac{d}{dt}\rho(t)&=&-i\left[H_{JC},\rho(t)\right]\nonumber\\
 \nonumber\\
 &+&\sum_{km=\pm1}\Gamma_{0m}\left(A_{0m}\rho(t)A^\dag_{0k}-A^\dag_{0k}A_{0m}\rho(t)\right)\nonumber\\
 \nonumber\\
 &+&\sum_{N'\ge1;k,l,m=\pm1}\Gamma_{0m}\left(A_{0m}\rho(t)A^\dag_{N'kl}-A^\dag_{N'kl}A_{0m}\rho(t)\right)\nonumber\\
 \nonumber\\
 &+&\sum_{N\ge1;k,m,n=\pm1}\Gamma_{Nmn}\left(A_{Nmn}\rho(t)A^\dag_{0k}-A^\dag_{0k}A_{Nmn}\rho(t)\right)\nonumber\\
 \nonumber\\
 &+&\sum_{N,N'\ge1;k,l,m,n=\pm1}\Gamma_{Nmn}\left(A_{Nmn}\rho(t)A^\dag_{N'kl}-A^\dag_{N'kl}A_{Nmn}\rho(t)\right)\nonumber\\
 \nonumber\\
 &+&\sum_{k,m=\pm1}\tilde{\Gamma}_{0m}\left(A_{0m}^\dag\rho(t)A_{0k}-A_{0k}A^\dag_{0m}\rho(t)\right)\nonumber\\
 \nonumber\\
 &+&\sum_{N'\ge1;k,l,m=\pm1}\tilde{\Gamma}_{0m}\left(A_{0m}^\dag\rho(t)A_{N'kl}-A_{N'kl}A^\dag_{0m}\rho(t)\right)\nonumber\\
 \nonumber\\
 &+&\sum_{N\ge1;k,m,n=\pm1}\tilde{\Gamma}_{Nmn}\left(A_{Nmn}^\dag\rho(t)A_{0k}-A_{0k}A^\dag_{Nmn}\rho(t)\right)\nonumber\\
 \nonumber\\
 &+&\sum_{N,N'\ge1;k,l,m,n=\pm1}\tilde{\Gamma}_{Nmn}\left(A_{Nmn}^\dag\rho(t)A_{N'kl}-A_{N'kl}A^\dag_{Nmn}\rho(t)\right)\nonumber\\
 \nonumber\\
 &+&\mathrm{ h.c.},
\end{eqnarray}
where $\Gamma_{0m}=\Gamma(E_{1,m}-E_0)$,
$\tilde{\Gamma}_{0m}=\Gamma(E_0-E_{1,m})$,
$\Gamma_{Nlm}=\Gamma(E_{N+1,l}-E_{N,m})$ and
$\tilde{\Gamma}_{Nlm}=\Gamma(E_{N,m}-E_{N+1,l})$ are complex
coefficients depending on the thermal spectral density of the
environment with which the atom-cavity system interacts, according
to:
\begin{eqnarray}\label{gammavsjomega}
 \Gamma(\omega)=\pi J(\omega)+i{\mathcal
 P}\int_{-\infty}^{+\infty}d\omega'\frac{J(\omega')}{\omega-\omega'},
\end{eqnarray}
where the thermal spectral density
$J(\omega)=\int_{-\infty}^{+\infty}\mathrm{e}^{-i\omega\tau}\left<E(\tau)E(0)\right>$
is the Fourier transform of the environment correlation function.
This quantity is in general equal
to\footnote{Eqs.~(\ref{gammavsjomega}) and~(\ref{Tdiversada0}) can
be obtained by generalizing the calculations given for the
two-level atom master equation in
Refs.~\cite{petruccionebook,gardiner}.}:
\begin{eqnarray}\label{Tdiversada0}
  J(\omega)=\left\{\begin{array}{lll}\left(n(\omega)+1\right)J_0(\omega)&,&\omega\ge0,\\
  \\
  n(\left|\omega\right|)J_0(\left|\omega\right|)&,&\omega<0,
  \end{array}
  \right.
\end{eqnarray}
where $J_0(\omega)$ is the zero-temperature spectral density,
defined for nonnegative frequencies, depending only on the density
of modes in the reservoir and on the distribution of the
system-reservoir coupling constants (i.e., the $g_k$ coefficients
in the model given in Eq.~(\ref{modello})), and
$n(\omega)=1/\left[\mathrm{exp}\left(\hbar\omega/k_BT\right)-1\right]$
is the average number of photons in a mode of frequency $\omega$,
for a general temperature $T$. The real part of $\Gamma(\omega)$
is equal to half the decay rate relative to the Bohr frequency
$\omega$, while the imaginary part is equal to the corresponding
Lamb shift~\cite{petruccionebook}.

Equation~(\ref{micros_model}) is the most general equation one can
give in accordance to the Born-Markov approximation and a RWA
performed on a timescale of the order of $\omega_0^{-1}$. Anyway,
Eq.~(\ref{micros_model}) is not in Lindblad form and in general it
might violate the complete positivity requirement for a master
equation~\cite{petruccionebook}: the possibility of casting the
Quasi-RWA master equation in a form suitable for physical
applications, i.e. Lindblad form, strongly depends on the specific
form of the spectral density of the environment and must be
evaluated in specific cases. We will see an important example in
the rest of the paper.

\section{Comparison: conditions for the validity of the phenomenological model}

In order to compare the phenomenological model to the Quasi-RWA
microscopic model given by Eq.~(\ref{micros_model}), it is
convenient to write Eq.~(\ref{phenME}) in the dressed states
basis. In the atom-cavity strong coupling regime, this procedure,
combined with a secular approximation, leads to the so called {\em
dressed state approximation} \cite{haroche,agar1,agar2}, which in
Ref.~\cite{Scala} we addressed as the main reason why the
phenomenological model is so successful in describing the
experimental situations.

To this aim let us write the cavity mode annihilation operator
with respect to the dressed states basis:
\begin{eqnarray}\label{annihilation_series}
 &&a=\sum_{N=0}^\infty\sqrt{N+1}\left|N\right>\left<N+1\right|\otimes
 \mathbb{I}_{\mathrm{at}}=\sum_{m=\pm1}A_{0m}+\sum_{N=1}^\infty\sum_{l,m=\pm1}A_{Nlm},
\end{eqnarray}
obtained by means of Eqs.~(\ref{eigenstates})
and~(\ref{groundeigenstate}) and of the relation
$\mathbb{I}_{\mathrm{at}}=\left(\left|g\right>\!\left<g\right|+
 \left|e\right>\!\left<e\right|\right)$.

Substituting Eq.~(\ref{annihilation_series}) and its hermitian
conjugate into Eq. (\ref{phenME}), one finally gets the
phenomenological master equation in the dressed states basis:
\begin{eqnarray}\label{dressed_phen_model}
 \frac{d}{dt}\rho(t)&=&-i\left[H_{JC},\rho(t)\right]\nonumber\\
 \nonumber\\
 &+&\gamma\left(n(\omega_0)+1\right)\left[\sum_{km=\pm1}\left(A_{0m}\rho(t)A^\dag_{0k}-\frac{1}{2}\left\{A^\dag_{0k}A_{0m},\rho(t)\right\}
 \right)\right.\nonumber\\
 \nonumber\\
 &+&\sum_{N'\ge1;k,l,m=\pm1}\left(A_{0m}\rho(t)A^\dag_{N'kl}-\frac{1}{2}\left\{A^\dag_{N'kl}A_{0m},\rho(t)\right\}\right)\nonumber\\
 \nonumber\\
 &+&\sum_{N\ge1;k,m,n=\pm1}\left(A_{Nmn}\rho(t)A^\dag_{0k}-\frac{1}{2}\left\{A^\dag_{0k}A_{Nmn},\rho(t)\right\}\right)\nonumber\\
 \nonumber\\
 &+&\left.\sum_{N,N'\ge1;k,l,m,n=\pm1}\left(A_{Nmn}\rho(t)A^\dag_{N'kl}-\frac{1}{2}\left\{A^\dag_{N'kl}A_{Nmn},\rho(t)\right\}\right)\right]\nonumber\\
 \nonumber\\
 &+&\gamma n(\omega_0)\left[\sum_{k,m=\pm1}\left(A_{0m}^\dag\rho(t)A_{0k}-\frac{1}{2}\left\{A_{0k}A^\dag_{0m},\rho(t)\right\}\right)\right.\nonumber\\
 \nonumber\\
 &+&\sum_{N'\ge1;k,l,m=\pm1}\left(A_{0m}^\dag\rho(t)A_{N'kl}-\frac{1}{2}\left\{A_{N'kl}A^\dag_{0m},\rho(t)\right\}\right)\nonumber\\
 \nonumber\\
 &+&\sum_{N\ge1;k,m,n=\pm1}\left(A_{Nmn}^\dag\rho(t)A_{0k}-\frac{1}{2}\left\{A_{0k}A^\dag_{Nmn},\rho(t)\right\}\right)\nonumber\\
 \nonumber\\
 &+&\!\left.\sum_{N,N'\ge1;k,l,m,n=\pm1}\!\left(A_{Nmn}^\dag\rho(t)A_{N'kl}-\frac{1}{2}\left\{A_{N'kl}A^\dag_{Nmn},\rho(t)\right\}\right)\right]\!.
\end{eqnarray}

The next step will be a comparison between the phenomenological
model in the form of Eq.~(\ref{dressed_phen_model}) and some
suitable forms of the microscopic model in
Eq.~(\ref{micros_model}).

From a first comparison between Eq.~(\ref{micros_model}) and
Eq.~(\ref{dressed_phen_model}), it is very easy to see that the
two models coincide when the temperature of the reservoir is zero
and its spectrum is flat, a condition strictly related to Markov
approximation, and under the hypothesis that all the imaginary
parts of the $\Gamma$ coefficients, i.e. all the Lamb shifts, are
negligible. Indeed under these assumptions all the $\Gamma$
coefficients in Eq.~(\ref{micros_model}) are equal to the zero
temperature spectral density $J_0$ (see Eq.~(\ref{Tdiversada0}))
and so is half the rate of cavity losses $\gamma/2$ in
Eq.~(\ref{dressed_phen_model}). Calling $\gamma/2$ the common
value of the $\Gamma$ coefficients, Eq.~(\ref{micros_model}) can
be rearranged in order to be exactly identical to
Eq.~(\ref{dressed_phen_model}) with $n(\omega_0)=0$.

This result provides a condition for the validity of the
phenomenological model at zero temperature: it is enough that the
spectral density is flat over the Bohr frequencies involved and
that all the renormalization terms arising from the imaginary
parts of the $\Gamma$ coefficients, see Eq.~(\ref{gammavsjomega}),
are negligible. The latter condition on the imaginary parts is
very often used in the literature, since it is assumed that the
Cauchy principal part of an integral of the spectral density is
very small compared to the real parts of the
coefficients~\cite{gardiner}. The assumption of flat spectrum,
instead, depends on the environment with which the atom-cavity
system is interacting, and on the initial state of the system:
indeed, since in view of the condition $T=0$ the energy of the
system cannot increase in time, one is allowed to truncate the
state space to a subspace with a finite number of excitations
equal to the maximum number of excitations in the initial state,
and it is in this space that all the decay rates must be the same.

Let us examine Eq.~(\ref{micros_model}) from another point of
view: the equation is written as a sum of many terms, most of
which are counter-rotating. Essentially the counter-rotating terms
are all the terms in the sums wherein the downward jump operator
is different from the hermitian conjugate of the upward jump
operator in the same term (for example the term
$A_{0+}\rho(t)A^\dag_{0-}$ in the first row). What we have
previously shown with our analysis is that, under the assumption
of zero temperature and flat spectrum, Eq.~(\ref{micros_model})
can be summed up and can be put in the form of Eq.~(\ref{oldME}).

Therefore an aspect emerging from our analysis is that the zero
temperature phenomenological model in Eq.~(\ref{oldME}) arises
microscopically from an interplay of many counter-rotating terms.
This point gives an interesting way of looking at the shift in the
frequency of the Rabi oscillations in the atom-cavity system
predicted by Eq.~(\ref{oldME})~\cite{Scala}. Indeed in the
phenomenological model, because of cavity losses, the Rabi
frequency is not exactly equal to $2\Omega$, but it is equal to
$2\Omega\sqrt{1-(\gamma/4\Omega)^2}$, i.e., the Rabi frequency is
shifted by a quantity which is second order in the ratio between
the decay rate $\gamma$ and the Rabi splitting $\Omega$. The shift
may be interpreted as a cooperative effect of the counter-rotating
terms in the dissipator, analogous to the Bloch-Siegert shift
appearing in the JC model when counter-rotating terms are added to
the atom-cavity interaction Hamiltonian~\cite{alleneberly}.

When the temperature of the reservoir is different from zero, the
comparison becomes more difficult to carry on. Let us assume that
the imaginary parts of the $\Gamma$ coefficients are all
negligible and that the zero temperature part $J_0(\omega)$ of the
spectral density of the environment is flat. According to
Eqs.~(\ref{gammavsjomega})~and~(\ref{Tdiversada0}), the real parts
of all the $\Gamma(\omega)$ factors, which are proportional to
$J(\omega)$, contain the quantities $n(\omega)+1$ for $\omega\ge0$
and $n(\omega)$ for $\omega<0$. Therefore in principle one should
claim (1) that neither can the $\Gamma$ coefficients be all equal
if $T\neq0$, nor can the $\tilde{\Gamma}$ coefficients, and (2)
that the phenomenological model in Eq.~(\ref{phenME}) can never be
microscopically justified.

Moreover in this case the microscopic Quasi-RWA master equation
may not be cast in the Lindblad form. This leads to some problems
in its use. On the one hand it may be very difficult to solve,
since in general there are not any standard techniques developed
for finding solutions of non-Lindblad master equations. On the
other hand there is the more fundamental problem that the general
form of Eq.~(\ref{micros_model}) may lead to unphysical results,
since it may violate complete positivity~\cite{petruccionebook}.
Hence, what we want to look for is a condition under which the
phenomenological model is at least approximately valid, so that
the problem of the cavity losses can be treated by means of
Eq.~(\ref{phenME}).

If during its time evolution the atom-cavity system has at most
$N$ excitations, the largest Bohr frequency involved in the
dynamics is $\omega_0+(\sqrt{N+1}+\sqrt{N})\Omega$ while the
smallest one is $\omega_0-(\sqrt{N+1}+\sqrt{N})\Omega$. Therefore
we have to check the magnitude of the difference between the rates
corresponding to the lowest and to the largest Bohr frequency
respectively. Since we are assuming that the quantity
$J_0(\omega)$ in Eq.~(\ref{Tdiversada0}) is flat, the difference
in the rates is all due to the difference in the photon
populations $n(\omega)$.

\begin{figure}
 \begin{center}
     \includegraphics{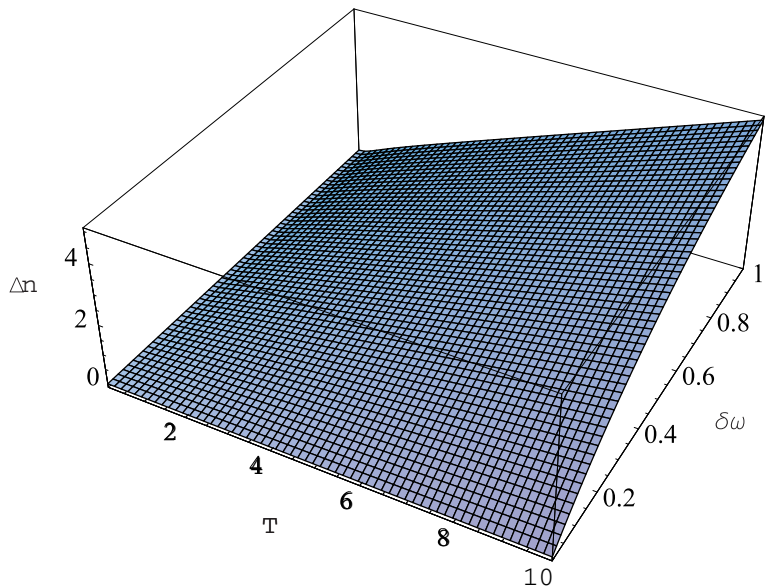}
    \caption{Difference $\Delta n$ in the average number of photons between a mode of frequency $\omega$ and a mode of frequency
    $\omega+\delta\omega$. The frequency $\omega$ is held fixed and the difference $\Delta n$ is shown as a function of the temperature
    $T$ (in units of $k_{\mathrm B}/\hbar\omega$) and of the frequency difference $\delta\omega$ (in units of $\omega$).}\label{validity}
 \end{center}
\end{figure}
Figure~\ref{validity} shows a plot of the difference in the
populations
$n(\omega)=1/\left[\mathrm{exp}\left(\hbar\omega/k_{\mathrm
B}T\right)-1\right]$ of two modes of the reservoir with
frequencies $\omega$ and $\omega+\delta\omega$ respectively, as a
function of the temperature $T$ and of the frequency difference
$\delta\omega$. From the plot one can argue that, for low
temperatures and not too large values of $\delta\omega$, the error
one makes assuming that all the decay rates in the master equation
are equal (and assuming the same for the excitation rates) can be
considered negligible. Since the $\delta\omega$'s in
Eq.~(\ref{micros_model}) are proportional to $\sqrt{N}$, the
negligibility of the difference in the rates imposes that the
number of excitations in the atom-cavity system cannot be too
large. Therefore we can conclude that, if the difference in the
rates can be neglected, the phenomenological model in
Eq.~(\ref{phenME}) can be considered as a valid model, provided
the temperature and the number of total excitations are not too
large.

\section{Discussion and Conclusive Remarks}
The analysis we have presented shows that, though the microscopic
and phenomenological models coincide when the spectrum of the
environment is flat, these two models are deeply different. The
discrepancy, which involves the predictions when the spectrum is
not flat and/or at $T\neq0$, is conceptual anyway, and relies on
the fact that the relevant physical mechanisms are different.

The point of view of our approach is the following: the general
formalism of master equations claims that all the decay and
decoherence channels involve the atom-cavity system as a whole,
and even in the cases wherein it is the cavity only which directly
decays also for the microscopic model, this may be seen as a
cooperative effect of many non-resonant terms describing the
coupling between populations and coherences relative to
atom-cavity dressed states.

A last point is worth mentioning. From our analysis we have
understood that the success of the phenomenological model relies
on the fact that the Quasi-RWA microscopic model reduces to it
when the spectrum of the environment is flat. The situation is
quite different when the spectrum is substantially non-flat. This
makes us interested in the possibility of studying the microscopic
model in the latter case, for which a non-Markovian theory may be
necessary. This is the scope of our future work.

\section*{Acknowledgements}

M.S. acknowledges financial support from CIMO for his stay in
Turku during the period January-April 2007 and wishes to thank all
the Quantum Optics Group members of the University of Turku for
their kind hospitality. S.M., J.P. and K.A.S. acknowledge
financial support from the Academy of Finland (projects 108699,
115982, 115682), the Magnus Ehrnrooth Foundation and the
V\"{a}is\"{a}l\"{a} Foundation.

\end{document}